# Emerging one-dimensionality from self-organization of electrons in NbSe$_3$

M. A. Valbuena[1,2]*, P. Chudzinski[3]*, S. Pons[1,4], S. Conejeros[5], P. Alemany[6], E. Canadell[7], H. Berger[1], E. Frantzeskakis[8], J. Avila[8], M. C. Asensio[8], T. Giamarchi[9] and M. Grioni[1]

[1]*Institute of Physics (IPHYS), Ecole Polytechnique Federale de Lausanne (EPFL), CH-1015 Lausanne, Switzerland*

[2] *Catalan Institute of Nanoscience and Nanotechnology (ICN2), CSIC and The Barcelona Institute of Science and Technology, Campus UAB, Bellaterra, 08193 Barcelona, Spain*

[3] *Institute for Theoretical Physics, Center for Extreme Matter and Emergent Phenomena, Utrecht University, Leuvenlaan 4, 3584 CE Utrecht, The Netherlands*

[4]*Laboratoire de Physique et d'Etudes des Materiaux,´ Ecole de Physique et Chimie Industrielles´ de la ville de Paris, ESPCI, PSL Research University; CNRS UMR8213; Sorbonne Universites,´ UPMC University, 10 rue Vauquelin, 75005 Paris, France*

[5] *Departamento de Quımica, Universidad Catolica del Norte, Av. Angamos 0610, Antofagasta` 124000, Chile.*

[6] *Departament de Ciencia de Materials i Quımica Fısica and Institut de Quımica Teorica i Computacional (IQTCUB), Universitat de Barcelona, Martı i Franques 1, 08028 Barcelona, Spain.´*

[7]*Institut de Ciencia de Materials de Barcelona (ICMAB-CSIC), Campus UAB, 08193 Bellaterra (Barcelona), Spain*

[8] *Synchrotron SOLEIL,Orme des Merisiers, Saint-Aubin-BP 48, 91192 Gif sur Yvette, France*

[9] *Department of Quantum Matter Physics, University of Geneva, 24 Quai Ernest-Ansermet, 1211 Geneva, Switzerland*

**Materials where the electron filling is close to commensurability provide one of the great challenges in materials science. Several proposals of unconventional orderings, where the electronic liquid self-organizes into components with distinct properties, were recently put for-**

**ward, in particular in cuprates[1–6] and pnictides[7,8] where electronic nematic orders have been observed. The electrons' self-organization is expected to yield complex intra and inter unit cell patterns, and a reduction of dimensionality. Nevertheless, an unambiguous experimental proof of such complex orders, namely the direct observation of distinct dispersions, is still missing. Here we report a Nano Angle Resolved Photoemission Spectroscopy (NanoARPES) study of NbSe$_3$, a material that has been considered a paradigm of charge order. The new data (Fig.1) invalidate the canonical picture of imperfect nesting[9] and reveals the emergence of a novel order. The electrons' self-organization uncovers the one-dimensional (1D) physics hidden in a material which naively should be the most 3D of all columnar chalcogenides.**

NbSe$_3$ undergoes two successive charge density waves (CDW) transitions[10] at $T_{C1}$ = 145 K and $T_{C2}$ = 59 K, affecting electrons on distinct columns[11–14] of its crystal structure shown in Fig.2(a). The charge ordering is not strong enough to remove the entire Fermi surface[10,15], and metallic conduction is observed even below $T_{CDW2}$. The nature of the gapless states has proven to be quite elusive, with several proposals[9,15–17] for the location of pockets within the Brillouin zone. In this paper we exploit NanoARPES with ultra-high lateral resolution and polarization control to probe these gapless states and to address the mechanism behind their formation (details on the experimental technique are available in Suppl. Mat.). NbSe$_3$ crystals grow in the form of very thin whiskers that assemble in bundles. The nanometric lateral resolution of the Antares beamline $k$-microscope proved crucial to select the best homogeneous areas in well-cleaved monodomain single crystals.

Figure 1 illustrates NanoARPES data collected with two photon polarizations along the $\Gamma Z$

(chains) direction in the low temperature phase ($T < 59$ K), where the sample experiences both CDW$_1$ and CDW$_2$. For s−polarization (Fig 1(a), perpendicular to the $b$ axis of the sample, we observe 3 parabolic bands, labelled A1$^+$, C2 and A3$^+$, near the Fermi energy ($E_F$), similar to previous ARPES investigations[9,18]. The effect of the CDWs is revealed by multiple gaps at $E_F$: in band C2 from CDW$_1$, and in bands A1$^+$ and A3$^+$ from CDW$_2$. From an analysis of spectra extracted from the energy-momentum map we estimate the gap values $\Delta_{C2} = \Delta_1 = 84$ meV and $\Delta_{A1+} = \Delta_{A3+} = \Delta_2 = 20$ meV, consistent with Ref.[9]. A gap in A3$^+$ at $k_{F2}$ ($E_{gap} \approx 2\Delta_1$ around $E_B = 0.18$ eV) and the apparent continuity of the folded bands C2 and A3$^+$ indicate an hybridization of these two bands driven by the CDW$_1$ potential. Data taken with p−polarization (Fig.1(d)) reveal different and unexpected features. Two linearly dispersing bands (labelled $A1^-$ and $A3^-$) now cross $E_F$ with no CDW-related gaps. These bands do not correspond to any of the spectral features observed with s−polarization for the same area of the single crystal. Such dramatic polarization dependence suggests that different types of carriers are present, and that they are separable by some selection rule.

As a first step to understand the experimental findings we performed advanced (GGA+U) DFT calculations. The inclusion of a local interaction term $U$ allows us to better differentiate Nb atoms with formally different oxidation states. The results of the calculation are shown in Fig.2(b), with further details in the Suppl. Mat. They confirm that the A1$^+$, A3$^+$ and C2 bands observed with s−polarization can be associated with crystal columns A and C. The calculated $k_F$ values are in good agreement with the measured data. The two bands labelled C are degenerate near $E_F$ as in the NanoARPES experiment where they appear as a single band, C2, implying a very small hybridization within a pair of C columns. Although the calculated bands crossing $E_F$ mainly derive from $d$ orbitals of a given Nb atom, the $d$ weight is never more than 50%, a rather different situation from e.g. transition metal oxides. This is a manifestation of the covalent character of the Nb-Se bonds and implies that the electron clouds are not locked on the Nb $d$ orbitals but potentially can be deformed when an extra potential from CDWs develops. The band dispersion has a quasi1D character, but the anisotropy is only moderate: the calculated bandwidth is $W_b$ = 2.5 eV along the b axis and up to 250 meV along perpendicular directions. The previously proposed band structures[9,16,17] hold also when local correlations are included, but $U$ causes vertical shifts of the bands labelled B and Se. Namely, band B is now located just above $E_F$, a result compatible with recent scanning tunneling microscopy measurements[14] where a significant spectral weight appears just above $E_F$. This may enhance inter-band backscattering, a process that controls the $CDW_2$ transition described below.

If the electrons could move within the crystal coherently in all directions, then the symmetry *planes* would be those of the $P2_1/m$ monoclinic space group of NbSe$_3$, parallel to the ac plane. Since the scattering plane in our NanoARPES experiment is perpendicular to these symmetry planes no definite selection rule should apply[19]. Identical dispersions should be observed for the two polarizations[20], in stark contrast with the measured data. The paradox is solved if correlations inhibit the coherent motion of electrons within the ac plane, endowing NbSe$_3$ with a strong onedimensional character. This implies that the standard nesting scenario

---

[1] The electrons are incoherent in the ac plane at high energies due to the hopping anisotropy (see Suppl. Mat. for calculation of an effective anisotropy)

for CDW formation, based on a warped 2D Fermi surface, must be abandoned and that a new description must be found.

To do so we first consider the bands C2 which, in our GGA+U calculations, are degenerate and nearly commensurate, and hence act as a doublet of 1D chains, prone to a Peierls distortion. Indeed the C2 dispersion is clearly backfolded, an effect of the CDW$_1$ with $\vec{q_1} = (0, 0.241b^*, 0)$, for which a charge distortion is known to be located mostly[11,12,14] on columns C. Its 1D character manifests itself in an anomalously large[21] ratio $\frac{\Delta_1}{T_{c1}} = 6.7$. The CDW$_1$ transition partitions the rest of the electronic liquid into 1D sub-systems[1] located on quartets of A+B columns (charge is predominantly on column A with some delocalization on column B). Each pair of A columns acts as a two-leg ladder system, where the competition of several internal interactions governs the physics[22–25]. An organization in the ac plane with a chessboard pattern (Fig.3(c)), which is a natural choice from the interactions viewpoint, allows to resolve the competition. Such pattern naturally yields the CDW$_2$ periodicity $q_2^\perp = (\frac{a^*}{2}, \frac{c^*}{2})$, which is hard to justify in a standard nesting scenario. Moreover this *ansatz* directly solves several experimental riddles: the band doubling into A1(3)$^+$ and A1(3)$^-$ and their tiny splitting, the number of electrons that remain ungapped, and the presence of gapless carriers close to the Γ point. For instance, dispersions A1(3)$^+$ and A1(3)⁻ correspond to newly formed bonding(+) / anti-bonding(-) states between adjacent quartets illustrated in Fig3 (b), that we shall call $\gamma^{\pm}_{1(3)}$.

---

[2] In the following we indicate by $\psi_{Ai}(r)$, the orbital that is a real space representation of an eigenstate corresponding to a DFT band Ai$^+$.

The above self-organization is governed by a dimerization along the c-axis. The driving force comes from a change in shape of the emerging eigen-orbitals ² $\psi_{\gamma\pm 1,3}(r)$ (of the states $\gamma^{\pm}_{1,3}$), which in turn changes the character of the electron-electron (e-e) interactions. For the emergent bonding state $\gamma^{+}_{1,3}$ the charge density of the constituent states $\psi_{A3}(r)$ (or $\psi_{A1}(r)$) extends towards column C (see Fig3(b)). While for $\psi_{A1,3}(r)$ there is only the long-range (Coulomb-type) interaction with CDW$_1$, which decays when the exchanged momentum increases, the emergent bonding states $\psi_{\gamma+1,3}(r)$ overlap more strongly with $\psi_C(r)$ and exchange processes become more likely. Hence the nature of the e-e interactions between carriers on column A and CDW$_1$ changes from long-range Coulomb type for $\psi_{A1,3}(r)$ to a more local (Hubbard) type for $\psi_{\gamma+1,3}(r)$. The scattering channel with large momentum exchange is accordingly enhanced. For the $\gamma^{+}_{3}$ state, the enhanced backscattering allows for a significant re-hybridization with a gapped band C2 (see Fig.1), an effect that has been noticed before[9]. The band structure of the $\gamma^{+}_{1,3}$ states is illustrated in Fig.3(a). The re-hybridization enhances e-e interactions with exchanged momentum close to ~$q_1$, as shown by arrows. Since ~$q_1$ is very close to $\vec{k}_{F1}+\vec{k}_{F3}$, the momentum exchanged in the inter-band backscattering processes, these processes are enhanced and can open a gap in the spectrum of the two-leg ladder (see Fig.3(a)). Thus, the emergent bonding states $\gamma^+$ are gapped and CDW$_2$, with $\vec{q}_2 = \vec{k}_{F1}+\vec{k}_{F3}$, has an intrinsically many-body origin.

Implications of our *ansatz* can be immediately tested. Firstly, since electrons move incoherently within the ac plane, an effective symmetry plane for the 1D system consisting of a quartet of A+B columns must: i) be perpendicular to the c-axis, ii) contain the b-axis, and iii) cross the middle of the quartet. The scattering plane $\sigma_l$ fulfils these conditions and thus the dipole selection rules emerge. Therefore the observed polarization dependence is a good indicator of the existence of a 1D regime. The bonding states $\gamma^+$ are by construction delocalized towards column C, hence they involve p-Se $\sigma$-hybridization on the short link (green line in Fig.2). These states are anti-symmetric with respect to the $\sigma_l$ plane (see Fig.3b) and hence are expected to appear in s–polarization, in agreement with the experiment. A $\pi$-bond component on the short bond, required to keep the total density close to the DFT solution, can be absorbed within the anti-bonding state $\gamma^-$. These states are symmetric (vs $\sigma_l$) and hence visible in p–polarization. Moreover, since they are localized mostly in the core of the quartet and interact weakly with CDW$_1$, they are expected to be gapless. These are the new states uncovered in p–polarization.

Another implication is that the gapless states $\bar{\gamma}^-_{1,3}$ should behave in accordance with the predictions of the 1D Tomonaga-Luttinger liquid (TLL) theory [Ref.26,27]. The momentum-integrated spectrum should exhibit a power-law dependence with a characteristic non-universal exponent $\alpha$. This behavior is compatible with the data with $\alpha = 0.24$ (see Fig.4(e)). Moreover, the $k$-resolved spectra should exhibit several modes (for a standard TLL there are two modes corresponding to spinon and holon, but for a ladder up to four modes are possible [27]) each with its own dispersion relation. Fits to the line shapes of energy- and momentum-distribution curves from our NanoARPES data in p–polarization expose that both the $A1^-$ and the $A3^-$ bands contain separate linear and parabolic dispersions, outlined with different symbols ("h" for holon and "s" for spinon) in Fig.4, that converge at $k_{F1,3}$. Note that these modes do not overlap with any features observed in the opposite polarization Fig.4(b). This strongly suggests a different behavior than in a Fermi liquid, in which only one velocity can appear. *All* signatures illustrated in Fig.4 are at least compatible with the TLL behavior.

In conclusion, NbSe$_3$ is the first material where electrons' self-organization into nano-sized emergent eigenstates with a strong 1D character is revealed by means of NanoARPES. This observation provide an insight into the CDW formation and self-differentiation mechanisms which are fully captured by our model resulting in two sets of states with distinct dispersions, one of which is gapped and the other one has TLL properties. The polarization dependence of NanoARPES allows these (orbitally distinct) states to be selectively probed. These results open the way to new experiments searching for the emergence of hidden 1D dimensional states in strongly correlated materials.

Acknowledgements

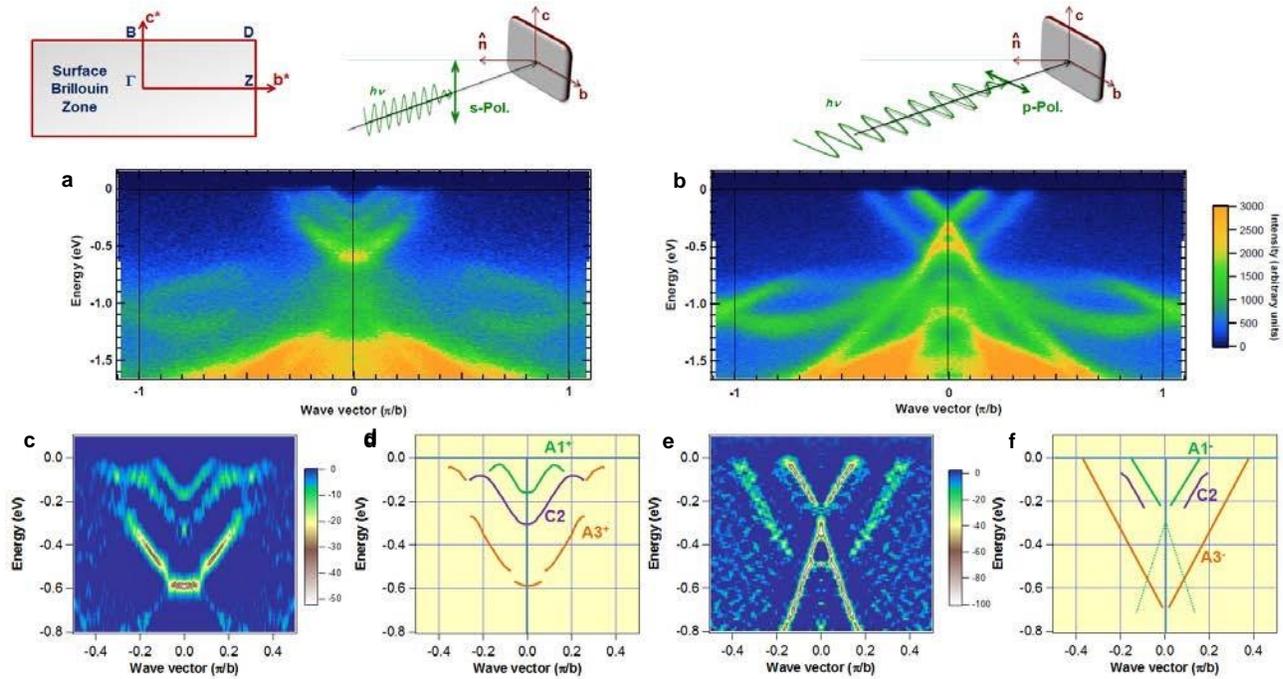

Figure 1: ARPES intensity maps taken in the $\Gamma b^*$ direction with photons of 100eV of two different polarizations. (a) s–polarization (b) ARPES p–polarization. Insets : experimental geometry and surface Brillouin zone. *s*-polarization is perpendicular to the long axis of NbSe$_3$ crystalline whiskers (i.e. perpendicular to the $b$ axis). *p*-polarization has a component along the long axis of the whisker (i.e. along the $b$ axis) and a component out of the plane. (c) Curvature procedure applied to the ARPES intensity map (a) (d) schematic of the band diagram recorded in (a) showing two parabolic bands associated to columns A (A1$^+$ & A3$^+$) and one parabolic band associated to columns C (C2). (e) Curvature procedure applied to the ARPES intensity map (b). Extrapolating the dispersions of the bands to $E_F$ yields $k_{F1} = 0.14\pi/b$; $k_{F2} = 0.22\pi/b$ and $k_{F3} = 0.37\pi/b$. (f) schematic of the band diagram recorded in (b) showing two new linear bands labeled $A1^-$ & $A3^-$ and a weak contribution of band C2.

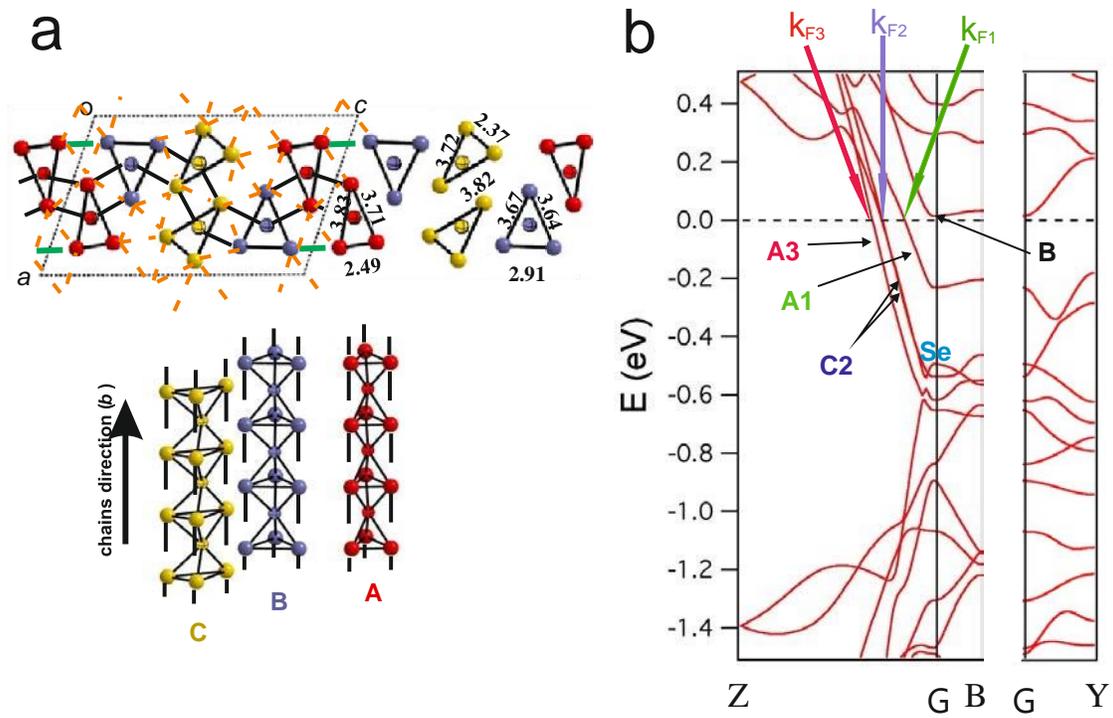

Figure 2: Crystal structure and DFT calculations. (a) Crystal structure of NbSe$_3$ built out of prismatic columns of three types, hereafter labeled column A-C. Bottom: View along the 1D chains (along the $\tilde{b}$ axis): Three types of columns are indicated in red (A), blue (B) and yellow (C).

Top: (*ac*) plane cross-section. Green (orange) dashed lines indicate strong (weak) Se...Se intercolumnar contacts. (b) Theoretical DFT GGA band calculations including finite $U = 4$ eV. Each band is indicated by the dominant orbital component. The bands can be associated with experimental counterparts presented in Fig.1 by comparison of the Fermi wave vectors. See supplementary material online for further calculation details.

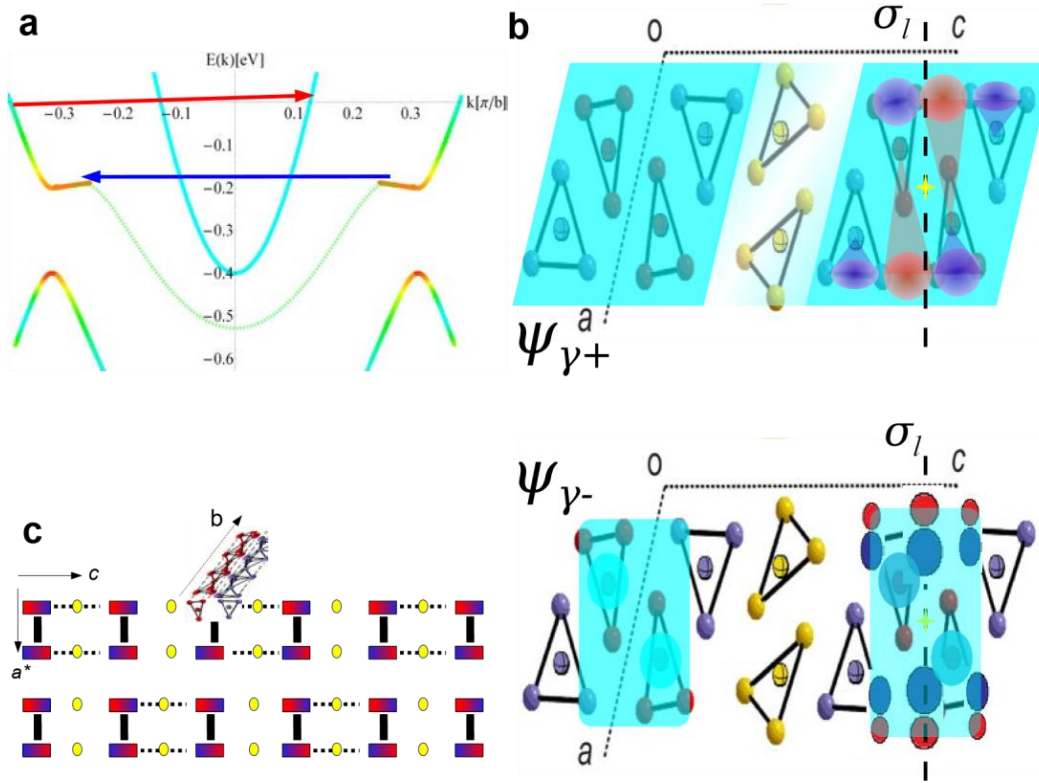

Figure 3: Electrons self-organization model. (a) Model band structure of re-hybridized $\gamma+$ states shows a proximity of an inter-band $A3^+ \rightarrow A1^+$ back-scattering (red-arrows) to a periodicity of CDW1 $q_1$ (blue arrow). The color scale encodes band curvature: when approaching red one approaches the van Hove singularity, where scattering processes are enhanced. (b) Bonding $\gamma_+$ (upper panel) and anti-bonding $\gamma_-$ (lower panel) eigen-orbitals emerging from a perpendicular dimerization between two neighboring unit cells of a basic crystal structure. Distinct wavefunctions are shown to have a different spatial extension. The amplitude is indicated by the strength of turquoise shading. Red and blue color indicates the phase of the involved *p*-orbitals. The orientation of a *p*-orbital determines the direction in which the resulting $\psi_\gamma$ orbital will be extended. The two wavefunctions are quite different, yielding *even* character on the bottom panel and *odd* character on the top panel (vs. the $\sigma_l$ plane indicated by the dashed line). (c) the cross section on a-c plane showing schematically the perpendicular dimerization and emergent self-organization. Yellow dots represent col. C where the "high energy physics", formation of $CDW_1$ takes place. Pairs of red-blue rectangles linked by thick black line are doublets of A+B columns. Bands $A1$ and $A3$ are bonding anti-bonding bonds formed on the black line. The perpendicular dimerization is indicated by the dashed black lines and it corresponds to bonding ($A1^+$ and $A3^+$) and anti-bonding ($A1^-$ and $A3^-$) combinations (with a tiny energy split).

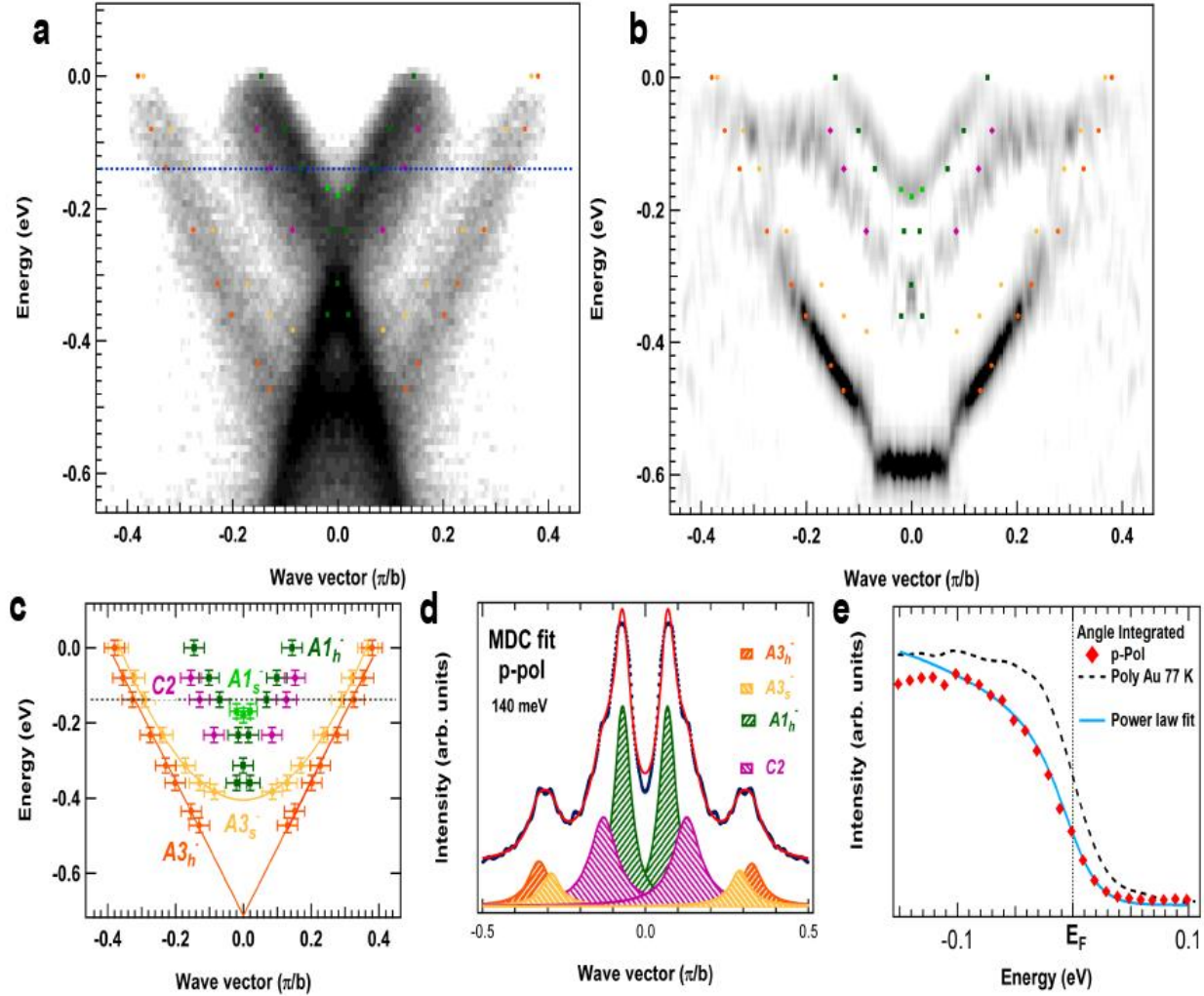

Figure 4: Signatures of 1D physics within the electronic states in NbSe$_3$:(a) Spectral intensity close to the Fermi level recorded with *p*polarization. Please note: observed extensions of $A1^-_h$ beyond the Dirac point at $E_{D1}$ = −0.25 eV), that were puzzling for DFT, are remarkably similar to the TLL-"shadow bands"[28]; (b) Comparison of observed data-points with the curvature image obtained with *s*–polarization. ; (c) Result of the fit of the spectral intensity measured with *p*-polarization as shown in panel(a). The fit evidences two component for each band $A$, $A1^-_s$ and $A1^-_h$, $A3^-_s$ and $A3^-_h$. Please note: the two double dispersions $A1^-,A3^-$ are geometrically similar – the ratios $E_{D1,3}/E_{B1,3}$) are equal (as expected for a two leg ladder). (d) example of MDC fit which has allowed us to find data points along the white dashed line in panel A (also indicated as dashed line in panel C); (e) Integrated intensity showing a power-law like vanishing of the density of states, compatible with a TLL behavior, fit following Ref.[29,30].

The work at Lausanne and Geneva is supported by the Swiss NSF. M.A.V. acknowledges financial support from Spanish MICIN Postdoctoral Mobility Program. SP acknowledges C'NANO Ile-de-France, DIM NanoK, for the support of the Nanospecs project. Work in Spain was supported by a MINECO (Grants FIS2015-64886-C5-4-P and CTQ2015-64579-C3-3-P), Generalitat de Catalunya (2014SGR301) and XRQTC. E.C. and M.A.V acknowledge support from MINECO through the Severo Ochoa Centers of Excellence Program under Grants SEV-2015-0496 and SEV2013-0295. SC gratefully acknowledges the Becas Chile program (CONICYT PAI/INDUSTRIA 72090772) for a doctoral grant at the Universitat de Barcelona. Synchrotron SOLEIL is supported by the Centre National de la Recherche Scientifique (CNRS) and the Commissariat a l'Energie` Atomique et aux energies Alternatives (CEA).

Author Contributions

H.B. grew the single crystal samples; M.A.V., S.P. and M.G. designed the experiment; M.A.V., S.P., J.A. and E.F. performed the experiments assisted by M.C.A.; M.A.V., and S.P. and J.A. analyzed the NanoARPES data; S.C, P.A. and E.C. performed the DFT calculations; P.C. and T.G. elaborated the theoretical model; P.C., S.P., E.C. and M.G. wrote the paper with comments from all co-authors.